# Large Solar Energetic Particle Events Associated with Filament Eruptions Outside of Active Regions

N. Gopalswamy

Solar Physics Laboratory, NASA Goddard Space Flight Center, Greenbelt, MD 20771

nat.gopalswamy@nasa.gov

P. Mäkelä[1], S. Akiyama[1] S. Yashiro[1], H. Xie[1], N. Thakur[1]

The Catholic University of America, Washington, DC 20064

S. W. Kahler

Air Force Research Laboratory, Albuquerque, NM 87117

¹also at NASA Goddard Space Flight Center, Greenbelt, MD 20771

## ABSTRACT

We report on four large filament eruptions (FEs) from solar cycles 23 and 24 that were associated with large solar energetic particle (SEP) events and interplanetary type II radio bursts. The post-eruption arcades corresponded to mostly C-class soft X-ray enhancements, but an M1.0 flare was associated with one event. However, the associated coronal mass ejections (CMEs) were fast (speeds ~1000 km/s) and appeared as halo CMEs in the coronagraph field of view. The interplanetary type II radio bursts occurred over a wide wavelength range indicating the existence of strong shocks throughout the inner heliosphere. No metric type II bursts were present in three events, indicating that the shocks formed beyond 2–3 Rs. In one case, there was a metric type II burst with low starting frequency indicating a shock formation height of ~2 Rs. The FE-associated SEP events did have softer spectra (spectral index >4) in the 10–100 MeV range, but there were other low-intensity SEP events with spectral indices ≥4. Some of these events are likely FE-SEP events, but were not classified so in the literature because they occurred close to active regions. Some were definitely associated with large active region flares, but the shock formation height was large. We definitely find a diminished role for flares and complex type III burst durations in these large SEP events. Fast CMEs and shock formation at larger distances from the Sun seem to be the primary characteristics of the FE-associated SEP events.

**Subject headings**: Sun: coronal mass ejections - Sun: filaments - Sun: flares - Sun: particle emission - Sun: radio radiation -   shock waves

# 1. Introduction

One of the key findings regarding solar energetic particle (SEP) events observed in the interplanetary medium has been the presence of fast coronal mass ejections (CMEs) indicating that fast CMEs drive MHD shocks that accelerate particles in the ambient corona (Kahler et al. 1978). Flares also accelerate SEPs that interact with the solar atmosphere and produce various electromagnetic bursts. Flare particles are also observed in the interplanetary medium as impulsive SEP events (see e.g., Reames 1999, 2013 for reviews). In eruptions occurring in active regions, it is often difficult to tell which mechanism is important because the eruptions involve both large flares and energetic CMEs. When CME-driven shocks arrive at an observing spacecraft, energetic storm particle (ESP) events are frequently observed, which represent the clearest evidence that CME-driven shocks accelerate particles (Reames 2012, Giacalone 2012). These events are detected after the shocks have propagated over one or two days.

There is another set of SEP events that primarily involve shock-acceleration—those associated with solar filament eruption events (Kahler et al. 1986). Flares in these events are extremely weak and are unlikely to be responsible for the large SEP events. Kahler et al. (1986) presented the case of the 1981 December 5 CME associated with a clear SEP event in which the observed proton energy range exceeded 50 MeV. The CME did not involve any active region, metric type II burst, or an impulsive-phase flare. However, the CME was associated with an interplanetary (IP) type II burst, implying shock formation in the IP medium. Another such SEP event on 1994 April 14 was associated with a post-eruption X-ray arcade (Kahler et al. 1998) although there were no accompanying white-light CME observations. While revisiting such events, Kahler (2000) suggested that the 1971 December 14 CME event (the first white-light CME ever detected (Tousey 1973)) might also be responsible for an SEP event and that it involved a filament eruption with little flare signature. Thus there were then a total of 3 known large SEP events originating from filament eruption events with very little flare signatures.

In this paper, we report on four large SEP events associated purely with filament-eruption CMEs that have complete data coverage from the Sun to the interplanetary medium. These events more than double the sample size of such events from three to seven. Furthermore, the complete set of available observations on solar sources, white-light CMEs, type II radio bursts, and interplanetary shocks makes it possible to derive unambiguous conclusions on these events. Two of the events were from solar cycle 23 and the other two from cycle 24. For the cycle-23 events, we have imagery from the Solar Heliospheric Observatory (SOHO, Domingo et al., 1995) and Yohkoh missions. For the cycle-24 events, we have the Solar Terrestrial Relations Observatory (STEREO, Howard et al. 2008) and the Solar Dynamics Observatory (SDO, Lemen et al. 2012) data in addition to the SOHO data. Radio spectral data are available from the Wind spacecraft (Acuña et al. 1995) for both cycles with the addition of STEREO data for cycle 24. Finally, SEP data are available from the SOHO, Wind, STEREO, and GOES missions.

## 2. Observations and Analysis

### 2.1 Overview of the observations

In this paper, we consider only large SEP events defined as those with a proton intensity in the >10 MeV GOES energy channel ≥10 pfu (pfu = particle flux unit; 1 pfu = 1 particle/$cm^2$.s.sr). In addition, we required that SEPs were detected in >50 MeV channel for a fair comparison with previous events reported by Kahler (2000). Table 1 provides an overview of the four events. All the four events have been listed in various papers involving statistical properties of SEP events or type II radio bursts (Gopalswamy et al. 2002; 2004; 2005; 2014; Gopalswamy 2012). The filament in the 2002 May 22 event was also studied extensively as it obscured the thermal emission from a neighboring flare (Gopalswamy and Yashiro 2013). The first two columns of Table 1 give the approximate onset time and the peak intensity (Ip) of the SEP events in the GOES >10 MeV channel. Column 3 gives the SEP intensity in the >50 MeV channel to check that these events were similar to the three events discussed by Kahler (2000). The solar source of the eruption is given in column 4 as the heliographic coordinates of the filament centroid. The size of the soft X-ray flare is given in column 5 either from the online Solar Geophysical Data or from our estimate based on the GOES soft X-ray time profile in the 1–8 Å channel. The first appearance time of the associated CME is given in column 6 as obtained from the CME catalog (cdaw.gsfc.nasa.gov; Yashiro et al. 2004; Gopalswamy et al. 2009) compiled from SOHO's Large Angle and Spectrometric Coronagraph (LASCO, Brueckner et al. 1995). The sky-plane speed of the CMEs ($V_{CME}$) from SOHO/LASCO observations is given in column 7 along with the space speeds in parentheses. The space speeds of the cycle-23 events were obtained from the cone model (Xie et al. 2004). For the last two events, space speeds were obtained using the combined SOHO-STEREO observations fit to the Thernisien (2011) graduated cylindrical shell model (Gopalswamy et al. 2014). The entry "H" in column 8 for CME widths (W) indicates that all CMEs were halos. Column 9 gives the average acceleration ($a_r$) in the coronagraph field of view. The wavelength range of the type II burst is given in column 10; this range is a good indicator of where the shock forms and how long it survives (Gopalswamy et al. 2005). Only one event had a type II burst starting from metric (m) wavelengths (starting frequency was 24 MHz). In all other cases, the type II burst started in the decameter-hectometric (DH) domain (below 13 MHz) as observed by the Radio and Plasma Wave Experiment (WAVES; Bougeret et al. 1995) on board the Wind spacecraft. In all cases, the radio emission continued down to kilometric (k) wavelengths, indicating that the shocks survived at least up to 1 AU. This was further confirmed by the shock detections at L1 by the SOHO MTOF proton monitor (Hovestadt et al. 1995); the shock arrival times at L1 are given in column 11. The final column (12) gives references where related information on these events can be found.

The composite plots of the four events in Fig.1 show the proton intensities in three GOES energy channels (>10 MeV, >50 MeV, and >100 MeV), CME height-time plots with the primary CME of interest along with other CMEs in the time intervals, and the Wind/WAVES dynamic spectra. The plots cover long time periods to show the full evolution of SEP intensity even after the shocks arrived at Earth (marked by the vertical lines in the radio dynamic spectra and the SEP intensity plots). The overview plots illustrate that these are Sun-to-Earth events. The shocks can be readily identified from the jumps in the local plasma frequency (fp) at Sun-Earth L1, where the Wind spacecraft is located. The shocks were also observed by the SOHO/MTOF instrument and listed in http://umtof.umd.edu/pm/FIGS.HTML. There was also some indication of the shocks in the particle intensity data as energetic storm particle (ESP) events. The 2002 May 22 event had the largest spike due to the ESP event, with an increase in the >10 MeV channel by an order magnitude. The 2000 September 12 and the 2011 November 26 events showed weak enhancements around the times of shock passage. On the other hand, the 2013 September 29 event showed a sudden drop in the SEP intensity, probably corresponding to the entrance of Earth into the associated ICME. A sudden impulse at Earth was found in all four events in the form of positive values of the Dst index (not shown, but available at http://wdc.kugi.kyoto-u.ac.jp/dstdir/index.html) shortly after the shock arrival times at L1. The impulse on 2002 May 23 was a sudden commencement, followed by a major storm (Dst = -109 nT). The sudden commencement on 2013 October 2 at 3:00 UT was followed by a moderate storm (Dst = -67 nT). The other two events did not produce significant magnetic storms.

In Figure 2, the proton intensity is plotted in several GOES differential energy channels that show the highest energy channel in which there is a discernible enhancement above the background. There was an enhancement above the background in the lowest 3 to 4 channels. The first and last events had increases in the 80–165 MeV channels, while the remaining two events had an increase only up to the 40–80 MeV energy channel. These observations are consistent with the plots in Fig.1, which show integral fluxes. The 2011 November 26 event shows a tiny enhancement in the >100 MeV (integral) channel, although no discernible increase was observed in the 80–165 MeV differential channel.

**2.2 SEP Spectra**

We also examined the proton intensity observed by the Energetic and Relativistic Nuclei and Electron (ERNE; Torsti et al. 1995) instrument on board SOHO. The intensities are generally consistent with the GOES observations. In order to get the energy spectra, we used the following ERNE energy channels: (1) 14.6–16.9 (15.8) MeV, (2) 16.9–30.0 (23.0) MeV, (3) 30.0–40.5 (35.2) MeV, (4) 40.5–47.0 (43.8) MeV, (5) 47.0–68.8 (58.2) MeV. The numbers in parentheses correspond to the effective energy of the channel. There were two other higher energy channels (effective energy 77.6 and 97.5 MeV) but the background level was too high to measure the SEP intensity, so we did not use them. Figure 3 shows the background-subtracted time-of-maximum differential energy spectra of the four SEP events in the energy range 10 MeV < E ≤ 100 MeV. In our estimation of the SEP spectra, we have excluded periods where the proton intensities peaked due to the ESP event. The power-law spectral index $\gamma$ ranges from 4.15 to 4.69, comparable to the 4.3 value derived for the 1981 December 5 SEP event by Kahler et al. (1986).

These values also lie in the high end of the approximate range of $2 < \gamma < 4.5$ for well-connected > 10 pfu SEP events of the later survey by Kahler (2001). Those differential spectral fits were determined from the two maximal event values of GOES E > 10 MeV and E > 60 MeV intensities.

**2.3 Solar Sources**

A variety of observations are available to pinpoint the solar sources of the CMEs. The filaments and flare ribbons were observed in H-alpha by ground-based observatories (Big Bear Solar Observatory in the United States and Hida Observatory in Japan). The photospheric magnetograms are from the SOHO Michelson Doppler Imager (MDI, Scherrer et al. 1995) and the SDO Helioseismic and Magnetic Imager (HMI, Scherrer et al. 2012). The post-eruption arcades were observed by SOHO Extreme Ultraviolet Imaging Telescope (EIT, Delaboudiniere et al. 1995) and SDO Atmospheric Imaging Assembly (AIA, Lemen et al. 2012). The soft X-ray light curves are from the GOES.

In the pre-eruption phase, the source regions were large-scale bipolar regions with filaments marking the polarity inversion lines of the magnetic structures. The filaments were best observed in H-alpha filtergrams, as shown in the top row of Fig. 4. The filament eruption was accompanied by a two-ribbon flare in each case, with the ribbons located on either side of the original position of the filament. The photospheric magnetograms show that the flare ribbons were located on opposite polarity regions, while the filaments were located on the polarity inversion line. The ribbons in the 2011 November 26 events were confined to the stronger magnetic field region, where the pre-eruption filament was very thin. There was another section of the filament that erupted from the southern end, but no ribbons were found around this section. The post-eruption arcades were also very extended, mostly in the north-south direction.

The bottom panels in Fig. 4 show that the soft X-ray flares associated with the four SEP events were generally weak, with the flare size ranging from C1.2 to M1.0 and were all listed in the Solar Geophysical Data. The M1.0 flare was associated with the 2000 September 12 event, which also had a weak impulsive phase: microwave emission was observed at the San Vito radio station of the Radio Solar Telescope Network (RSTN) up to 15.4 GHz. The microwave intensity was ~46 solar flux units (sfu; 1sfu = $10^{-22}$ Wm$^{-2}$Hz$^{-1}$). There was no hard X-ray data from Yohkoh/HXT for this event. None of the other flares had impulsive phase microwave emission.

## 2.4 White light Observations and CME Kinematics

The eruptive filaments can be traced as cores of the associated white-light CMEs (see Fig. 5). The CMEs were fast and wide in all four cases. Figure 5 shows snapshots of the CMEs within the field of view of the outer telescope (C3) of LASCO on board SOHO. The images were taken about 2–3 hours after the first appearances. The morphologies of the prominence cores are consistent with the polarity inversion lines and the post-eruption arcades in Fig. 4. The CME speeds in the sky plane, as measured from the LASCO images, range from 933 km/s to 1550 km/s. For the halo CMEs in cycle 23, a cone model has been used to correct for projection effects (Xie et al. 2004, Gopalswamy et al. 2010). The deprojected space speeds were higher: 976 km/s to 1946 km/s. For the last two events, Gopalswamy et al. (2014) obtained the space speed from the Thernisien (2011) flux rope model by combining the SOHO and STEREO observations. The space speeds were not too different from the cone model values (1187 km/s vs. 976 km/s and 1864 km/s vs. 1543 km/s). Since the CMEs were all halos, they are expected to be wide (also clear from Fig. 5). Therefore, all CMEs in Table 1 meet the criterion of fast ($\geq$ 900 km/s) and wide ($\geq$ 60º) CMEs to be driving shocks that accelerate particles. The speeds of the CMEs considered in this paper were somewhat higher than the sky plane speed of 840 km/s for the 1981 December 5 CME reported by Kahler et al. (1986).

**2.5 Prominence Speeds**

We were able to measure the height-time history of three of the four eruptive filaments using EUV images in the early phases and LASCO images in the later phases when the filaments became white light cores inside CMEs. In the case of the 2011 November 26 eruption, the eruptive filament was not clearly visible to make the height-time measurements, so we used the initial location of the filament in H-alpha and followed the filament as the prominence core in white light. We also used the CME leading edge measurements available at the SOHO/LASCO CME catalog.

The height-time histories of the four events are shown in Figure 6. The speeds of the prominences in the three events were roughly ~275 km/s by the time they reached about 0.5 Rs above the surface. This value is similar to the prominence speed of 305 km/s reported by Kahler et al. (1986). The height-time plots show that the separations between the leading edges (LE) of the CMEs and the corresponding prominence cores steadily increased because of the different speeds, which is well known (Gopalswamy et al. 2003). The core-LE separation was the smallest for the November 26 event and the final speeds differed only by a factor of 1.2. The filaments accelerated to much higher speeds when they became eruptive prominence cores of the white-light CMEs (by more than a factor of 2—see Table 2). However, the CME leading edge speeds were higher by a factor of up to 4. The initial average acceleration of the filaments obtained from the EUV measurements are at the higher end of the values obtained for the general population eruptive prominences (Gopalswamy 2015).

**2.6 Comparison with Flare Acceleration**

Soft X-ray flares were well observed in all four events, so we estimated the initial acceleration of CMEs from the flare rise time and the average CME speed in the LASCO field of view (FOV) assuming that the CMEs finished accelerating by the time they reached the LASCO FOV (Zhang and Dere 2006; Gopalswamy et al. 2012). The average initial acceleration of the CMEs was in the range from 0.28 km/s$^2$ to 0.80 km/s$^2$, much smaller than that of CMEs producing GLE events (average ~2 km/s$^2$ see Gopalswamy et al. 2012). The initial accelerations were also generally lower than that (~1 km/s$^2$) of limb CMEs associated with soft X-ray flares of importance $\geq$C3.0. On the other hand, the accelerations were significantly higher than those of the general population of filament eruption events ($\leq$150 m/s$^2$ – see Gopalswamy 2015). The accelerations of the eruptive prominences obtained from the height-time plots in Fig. 6 are also lower than those derived from the flare rise time and CME leading edge speed in the LASCO FOV.

**2.7 Radio Observations**

Type II radio bursts in the metric and longer wavelengths are of primary interest because they are indicative of shock-accelerated particles. As noted in Table 1, a type II burst was associated with each of the SEP events. A metric type II burst was observed only in the 2000 September 12 event. Type II bursts were observed in the DH and kilometric domains in all four events. Radio observations were not available in the DH domain for the previous EP-associated SEP events, so the shock formation height could not be determined. The highest frequency of the ISEE-3 radio instrument was 1.98 MHz (Knoll et al. 1978), compared to 13.8 MHz in Wind/WAVES and 16 MHz in STEREO/WAVES. The frequency range 2–14 MHz corresponds to plasma frequencies in the 3–10 Rs heliocentric range and hence provides radio observations filling the gap between metric and kilometric observations. Kahler et al. (1986) reported that the type II burst during the 1981 December 5 event was barely discernible at 1.98 MHz. In the present events, the type II bursts were observed from 13 MHz and below.

Figure 7 shows composite dynamic spectra constructed from ground based RSTN observations and Wind/WAVES observations. All the events show fundamental harmonic structure at frequencies below 1 MHz, except the May 22 event, which had an accompanying continuum and a radio data gap. A metric type II burst was observed only during the September 12 event. The metric type II started at ~48 MHz. There was a weak type II emission between 10 and 5 MHz in the DH domain. Finally the type II appeared with fundamental-harmonic structure at frequencies below 2 MHz. A line-up of the type II features at various wavelengths suggests that the metric type II burst is the harmonic component, which means the type II burst started at a plasma frequency of 24 MHz.

The fundamental component at frequencies below about 2 MHz intensified in all four cases. In all events, the type II bursts were weak and of narrow band in the DH domain (the sensitivity of the Wind/WAVES antenna in the 1–14 MHz range is lower than that at lower frequencies), but intensified around 1 MHz. The starting frequency of the November 26 and September 29 events were at 10 MHz with no metric component. The only case with much lower starting frequency is the May 22 event, in which the type II burst started at 0.5 MHz, again with no metric component. In this event, there was an ongoing intense continuum emission between 5 MHz and 0.4 MHz, so it is difficult to conclude on the type II emission in this range. However, the starting of the type II burst at 0.5 MHz is quite clear, so we think there was no type II emission at higher frequencies. The circumstances of the May 22 event are somewhat different, which we discuss later (see below).

The starting frequency of a type II burst is of particular interest because it is indicative of the height of shock formation ahead of the CME. By studying a large number of type II bursts associated with EUV waves or white-light CMEs, Gopalswamy et al. (2013) were able to determine the heights of shock formation. They found that the starting frequency ($f$) of a type II is smaller when the shock forms at a greater heliocentric distance according to the empirical relation, $f = 308.17r^{-3.78} - 0.14$, where $f$ is in MHz and $r$ is the heliocentric distance in solar radii. For the 2000 September 12 event, $f = 24$ MHz, so the shock formation height was $r = 1.97$ Rs, corresponding to the outer corona. This is consistent with the estimate of ~1.92 Rs made from the flare acceleration method (Mäkelä et al. 2015). For the November 26 and September 29 events, the shock formation heights were at 2.5 Rs because the starting frequencies of the type II bursts were 10 MHz. For the May 22 event, $f=0.5$ MHz, so we get $r=5.1$ Rs. This distance is well below the measured CME height (8.48 Rs) in the LASCO FOV. If we use the cone model result, the space speed is larger than the sky plane speed by a factor of 1.15, and we get a CME height of 9.75 Rs. This is almost twice the height implied by the empirical relation. The discrepancy can be readily understood from Fig. 5, which shows that the May 22 CME is entering into the region depleted by the preceding CME. By comparing the actual CME height at the type II start with the expected height from statistics, we see that the plasma frequency was lower by a factor of ~1.9. This means the density was depleted by a factor ~3.7. An alternative explanation would be that the radio emission originated from the flanks of the shock. However, the Alfven speed at the flanks is expected to be higher because the plasma level corresponds to the deceasing side of the Alfven speed profile (Mann et al. 1999; Gopalswamy et al. 2001). We conclude that the severe deviation from the empirical formula is due to the large depletion caused by the preceding CME. This is further supported by the fact that the type II burst at later times (beyond 14 UT after a data gap of 8 hr) was observed at a frequency of 0.2 MHz. At this time, the CME was expected to be at a heliocentric distance of >50 Rs. The type II burst frequency at this time appears normal, which means the CME merged with the preceding one and the resultant CME moved through the normal interplanetary medium (no depletion).

Table 3 shows the shock speeds derived from the drift rates of type II bursts, and CME space speeds are given for comparison. The drift rates were obtained in the 1–13 MHz and <1 MHz bands. These roughly correspond to the outer corona and IP medium, respectively (see e.g. Gopalswamy 2011). In deriving the density scale heights, we assumed that the density $n$ varied as $n_o r^{-\alpha}$, where $r$ is the heliocentric distance, $\alpha$ is the power law index and $n_o$ is the coronal base density. This form gives a simple estimate of the scale height as $r/\alpha$. We use the heliocentric distance of the CME nose for $r$. For cycle-23 events, we use the cone model to estimate the projection effects. For cycle-24 events, the heliocentric distance is obtained by the forward-fitting technique to the STEREO and SOHO coronagraph images. In the IP medium, $\alpha = 2$, whereas it is higher in the corona. Table 3 gives $\alpha$ values that yielded shock speeds close to the CME speeds. In the 1–13 MHz band, $\alpha = 4$ was suitable except for the September 29 event, where $\alpha = 2$ was suitable. Type II bursts were fragmented especially in the 1–13 MHz band, so some of the estimates are uncertain. Similarly, $\alpha = 3$ was suitable at 0.6 MHz, which corresponds to the transition from the outer corona to IP medium. The main purpose of this analysis is to show that both the shock speed and the CME speed obtained from independent techniques indicate that the CME speeds were very high, suggesting that the SEPs are consistent with shock acceleration.

There were intense type III bursts in all events, but only the cycle-23 events were of long duration at 1 MHz (35 min and 40 min for the September 12 and May 22 events). The November 26 event had a duration of only15 min, while the September 29 event had the shortest duration (~10 min). The 14 MHz type III burst was either non-existent or of extremely short duration. One interesting point is that the onset of the type II burst preceded that of the type III bursts in all cases, except in the May 22 event. It has been shown before that the existence of a long duration type III burst is not a sufficient condition for SEP association (Gopalswamy and Mäkelä 2010). The two cycle-24 SEP events show that the long-duration type III burst is also not necessary. The soft X-ray flares are the weakest in the cycle-24 events and have shorter-duration type III bursts, suggesting that they may not be related to the CME-driven shock.

## 3. Discussion

We analyzed four FE-SEP events from solar cycles 23 and 24 for which complete information was available on the filament eruptions and the associated CMEs. The filament eruptions occurred in quiescent regions with the two-ribbon flares and post-eruption arcades characterizing these events. The soft X-ray enhancements were in the range C1.2 to M1.0 with a weak impulsive phase in the M1.0 event. The eruptive filaments typically had higher accelerations compared to the general population (Gopalswamy 2015). The eruptive filaments eventually became the cores inside the white-light CMEs and were trailing the CME leading edge as far as we could track them within the coronagraph FOV. The associated CMEs were all fast (1000 to 2000 km/s) and full halos, indicating that they were capable of driving fast mode MHD shocks needed to energize the particles.

### 3.1 Initial CME Acceleration

The CME leading-edge accelerations derived from the flare rise times and CME speeds in the outer corona were within the range of values obtained for non-GLE large SEP events (Mäkelä et al. 2015). Figure 8 shows the distribution of CME accelerations for 59 large SEP events from cycles 23 and 24 (1997 November to 2014 January). The four FE-SEP event values are shown by arrows. Note that all the four CME accelerations are below the mean and median values of the distribution. However, two of the CME accelerations in FE-SEP events (September 12 and May 22) are in the same bin as the most probable value of the distribution. The September 29 event is in the lowest bin. Only the 2011 November 26 event is the least impulsive event with an acceleration of 0.28 km/s$^2$, which is lower than the lowest bin of the distribution.

### 3.2 Proton Intensities in the FE-SEP Events

The peak >10 MeV proton intensities ranged from 80 to 820 pfu in the four FE-SEP events studied in this paper. It must be noted that there are many large SEP events with >10 MeV proton intensities smaller than those of the FE-SEP events. The 1981 December 5 FE-SEP event (Kahler et al. 1986) was not found in the NOAA proton events list compiled since 1976. However, we examined the ISEE-3 plots available in the Solar Geophysical Data (comprehensive report #475) and found that the event had 0.1, 0.017, and 4x10$^{-4}$ particles per (cm$^2$.s.sr.MeV) in the 13.7–25.2 MeV, 20–40 MeV, and 40–80 MeV channels, respectively. By fitting an approximate spectrum to these values and assuming that there were no significant contribution from particles with energies >100 MeV, we estimated a >10 MeV flux of 14.5 pfu for the 1981 December 5 FE-SEP event. Thus the FE-SEP events in the present study are 1-2 orders of magnitude larger than the 1981 December 5 event.

### 3.3 Time-of-maximum Spectra

The spectral index γ at the peak SEP intensity in the 10–100 MeV energy range was generally high (4.15 to 4.69) matching that of the 1981 December 5 event (Kahler et al. 1986) and in the upper range of the two-point spectral indices of Kahler (2001). We compared the spectra of FE-SEP events with those of two subsets of SEP events: (i) large SEP events associated with weaker soft X-ray flares (C-class or smaller), and (ii) large SEP events with low proton intensity (<30 pfu in the >10 MeV GOES channel). We now examine these two subsets in more detail.

### 3.3.1 SEP Events Associated with C-class Flares

The large SEP events associated with C-class flares are shown in Table 4 (excluding the three FE-SEP events associated with C-class flares). The soft X-ray importance was in the range C3.7 to C9.7. The SEP intensities (Ip) were in the range 14-55 pfu, smaller than those of FE-SEP events. The spectral index $\gamma$ was obtained by fitting a curve of the form $dN/dE = kE^{-\gamma}$ to the ERNE measurements in the 10–100 MeV range as in Fig. 2. The sky-plane speeds ($V_{CME}$) and widths ($W$) of the CMEs were obtained from the SOHO/LASCO CME catalog. The eruption locations were all in the western hemisphere similar to the FE-SEP events. The association of a metric type II burst (y = Yes, n = No) burst and/or a DH type II burst (Y=Yes, N=No) is indicated in the last column of Table 4. All events had DH type II bursts and only two (2000 April 4 and 2010 August 14) had both metric and DH type II bursts (yY events). The metric type II burst during the 2000 April 4 event was extremely weak. The fundamental component was reported to be at frequencies below 80 MHz. Our examination of the dynamic spectrum reported in Kahler et al. (2007) indicated that the starting frequency was around 40 MHz. The metric type II burst during the 2010 August 14 event was also brief but had clear fundamental harmonic structure in the e-CALLISTO dynamic spectrum obtained at the Gauribidanur Radio Observatory near Bangalore in India (http://soleil.i4ds.ch/solarradio/qkl/2010/08/14/GAURI_20100814_094458_59.fit.gz.png). The fundamental component started at 60 MHz, indicating larger CME height at type II onset. The empirical formula used in section 3 suggests that the shock heights were at 1.72 and 1.54 Rs, respectively for the 2000 April 4 and 2010 August 14 events. The DH type II bursts were intense and extended to frequencies below 1 MHz, except for the 2010 August 14 event. In this event, the DH type II burst was extremely narrow banded and ended at ~3 MHz. Since most of the events in Table 4 had only DH type II bursts, we infer that the large CME height is an important characteristic of these events, a property shared by the FE-SEP events. The main difference between the events in Table 4 and the FE-SEP events is the higher >10 MeV intensities of the latter.

Figure 9 shows the SEP spectra of the five events in Table 4. The spectral index ranged from 3.25 to 4.76, overlapping with that of the FE-SEP events. Note that two events had spectral index >4 and one had spectral index close to 4 (2000 October 25). The spectral indices of the last two events in Table 4 were well below those of the FE-SEP events.

The 2000 April 4 event had the largest $\gamma$, even larger than the largest value (4.69) for the FE-SEP events. Puzzled by this, we examined the source region of the 2000 April 4 event in more detail. We found that the eruption was ~10º to the east of NOAA AR 8933. Therefore, this event should have been classified as a purely filament eruption event. This is further demonstrated in Fig. 10 using H-alpha images and magnetograms from the Big Bear Solar Observatory (BBSO) combined with SOHO/EIT images. A large north-south filament was located to the east of NOAA AR 8933 as shown the H-alpha image taken at 16:34 UT on 2004 April 3 (Fig. 10a). The filament centroid was at N14W55, whereas the active region was located at N16W66. The BBSO videomagnetogram (VMG) taken around the same time as the H-alpha image shows that the filament was located in a weak field region (Fig. 10b). The negative polarity of the weak field region appears to be an extension of the AR polarity. A post eruption arcade observed by SOHO/EIT at 195 Å on April 4 at 17:36 UT (Fig. 10c) was also clearly to the east of the active region. Finally, the H-alpha image taken after eruption shows that the two-ribbon structure had formed clearly to the east of the active region. Thus, the 2004 April event is indeed a filament eruption event, similar to the other FE-SEP events considered in this paper.

The source region of the 2004 April 11 event also was somewhat similar to the 2000 April 4 event: there was a circular filament mostly outside AR 0588. This event was studied in great detail to show the relation between type III bursts, type II bursts, and SEP events (Gopalswamy and Mäkelä 2010). The associated flare had some impulsive phase emission as indicated by the microwave emission at frequencies up to 15400 MHz (intensity ~220 sfu with peak value of 920 sfu at 2695 MHz). This event was listed as a disappearing solar filament (DSF) associated with AR 0588. Thus the 2004 April 11 event is also likely to be a filament eruption event, although it is not as clear as the 2004 April 4 event.

The 2000 October 25 event was also a filament eruption event, but the filament was located between two large active regions 9201 and 9199 in the western hemisphere. Yohkoh soft X-ray images show a large arcade formed between these two ARs. This event is also likely to be a FE type event. The spectral index was also relatively high (3.82). The remaining two events in Table 4 were from active regions and the spectral indices were smaller.

### 3.3.2 Low-intensity SEP Events

Table 5 lists 9 large SEP events with intensity ≤30 pfu. In fact, there were about 50 events with such low intensity, but we have listed only those that were well connected (similar to the FE-SEP events). We also eliminated a few events in which the determination of the peak intensity was difficult due to adjacent events. In two cases, there was ERNE data gap, so we were not able to compute the spectrum. Table 5 shows that the low-Ip events were all associated with larger flares (the lowest was an M1.5 flare on 2001 September 15; the CME speed was also low – only 478 km/s). All events were also associated with impulsive microwave emission at frequencies up to 15400 MHz. The type II association is remarkably uniform in these events: all but one had both metric and DH type II bursts (yY events) indicating shock formation closer to the solar surface in these events. The lone exception was the 2002 March 16 event. The eruption started towards the end of March 15 from close to the disk center and was associated with a halo CME. There was also a weak impulsive microwave emission up to 15400 MHz (62 sfu). There was a type IV burst in the metric domain, but no type II burst. The type II burst started around 14 MHz and was rather weak in the Wind/WAVES dynamic spectrum. Clearly the shock formation was at a larger height, similar to the FE-SEP events, but this event is from an active region (AR 9866). Interestingly, the spectral index for this event was 4.52, very similar to the FE-SEP events. The common aspect between this nY event and FE-SEP events seems to be the shock formation at larger heights. There was one other event (2002 August 14) with a large spectral index (4.58). The event was associated with a metric type II burst, but the starting frequency was only 50 MHz, again indicating a shock formation height of ~1.62 Rs as derived from the empirical formula of section 3. The spectral indices of the remaining events were in the range 1.67 to 3.82 indicating harder spectra than in the FE-SEP events. In these events, the starting frequencies of metric type II bursts were generally higher, indicating shock formation much closer to the Sun.

**3.4 Interplanetary Type II Bursts during the FE-SEP Events**

In the FE-SEP event (1981 December 5) reported by Kahler et al. (1986), the interplanetary type II burst was barely discernible, in contrast to the events in Table 1, all of which show intense type II bursts in the IP medium that could be tracked for long distances from the Sun. The 1-AU shocks were also readily identified. One important outcome of this study is that the lack of a metric type II burst is not necessarily a significant characteristic. However, shock formations at heliocentric distances close to or larger than ~2 Rs are indicated in the FE-SEP events. The Alfven speed profiles typically peak around 3 Rs, so the shock formation happens at a distance where the Alfven speed is still increasing with height. The large CME speeds measured from white-light imagery and type II radio observations make them sufficiently super-Alfvenic to accelerate particles. The continued acceleration of CMEs in the FE-SEP events make them more super-Alfvenic when they reach the declining portion of the Alfven speed profile as indicated by the intense and long-lasting type II bursts in the IP medium. Mäkelä et al. (2015) reported that the shock formation height was >3 Rs in a few SEP events with metric type II bursts. One has to examine the property of the ambient corona and the CME acceleration profile in detail for these cases, including consideration given to the possible type II formation at the shock flanks.

## 4. Summary and Conclusions

We presented four clear cases of large SEP events caused by CME-driven shocks involving filament eruptions outside of active regions. We were able to obtain a complete picture of these eruptive events because detailed solar imagery and radio dynamic spectra were available from multiple instruments from space and ground. These events form an important subset of SEP events in which the SEPs are definitely accelerated by shocks rather than by flare reconnection. In this respect (shock acceleration) there is no difference between CMEs from active regions or quiescent filament regions. The main difference is the weakness of the flares in H-alpha and soft X-rays, indicating that the filament eruption events have no or weak impulsive phases. All the events involved relatively strong shocks that survived all the way to Earth. Shocks were indicated by type II bursts near the Sun and IP medium. The shock formation height was in the range 2–3 Rs for three events and > 5 Rs in one case. In the case of previous filament-eruption SEP events, there were no radio observations above 2 MHz, so the shock formation height could not be determined.

We confirmed that the spectral index in the 10–100 MeV energy range was large, typically >4 for the FE-SEP events. We also examined large SEP events associated with C-class flares. There were five such events. Three of these five events had large spectral indices (3.82 to 4.76). On further examination we found that these three events were filament eruption events and were incorrectly associated with nearby active regions. These events also had fast CMEs and DH type II bursts with large shock formation heights. We also examined low-intensity (≤30 pfu) SEP events from longitudes similar to the FE-SEP events. Even though these events were all associated with larger flares (>M2) from active regions, two of them had spectral index >4, similar to the FE-SEP events. One event did not have a metric type II burst and the other had a metric type II with low starting frequency indicating that the shock formation height was large, which seems to be a common property of all the high-spectral index events (FE-SEP events or not). The duration of complex type III bursts were also not significant in the FE-SEP events. On the other hand, the availability of a fast CME seems to be the primary characteristic of these CMEs as evidenced by the shock-related radio emission in the IP medium and the arrival of shocks at 1 AU.

**Acknowledgements**

We thank the Big Bear Solar Observatory and the Hida Observatory for making their H-alpha data available on line. We thank NOAA/NGDC for making the GOES proton data available. This work benefitted greatly from the open data policy of NASA. STEREO is a mission in NASA's Solar Terrestrial Probes program. SOHO is a project of international collaboration between ESA and NASA. The work of NG, SY, SA, and NT was supported by NASA/LWS program. PM was partially supported by NSF grant AGS-1358274 and NASA grant NNX15AB77G. HX was partially supported by NASA grant NNX15AB70G.

between ESA and NASA. The work of NG, SY, SA, and NT was supported by NASA/LWS program. PM was partially supported by NSF grant AGS-1358274 and NASA grant NNX15AB77G. HX was partially supported by NASA grant NNX15AB70G.

Table 1. Properties of large SEP events associated with prominence-eruption CMEs

| SEP UT | Ip (10 pfu[a]) | Ip (50 pfu[b]) | Source | Flare size[c] | CME UT[d] | $V_{CME}$ km/s[e] | W | $a_r$ ms$^{-2}$ | Type II[f] | IP shock | Ref.[g] |
|---|---|---|---|---|---|---|---|---|---|---|---|
| 2000/09/12 13:00 | 320 | 1.5 | S17W09 | M1.0 | 11:54 | 1550 (1946) | H | 58.2 | mk | 9/15 04:28 | 1-4 |
| 2002/05/22 06:00 | 820 | 0.4 | S30W34 | C5.0 | 03:50 | 1494 (1718) | H | -10.4 | Dk | 05/23 10:40 | 2-5 |
| 2011/11/26 11:25 | 80 | 0.4 | N27W49 | C1.2 | 07:12 | 933 (976) | H | +9.0 | Dk | 11/28 21:27 | 6 |
| 2013/09/29 23:50 | 180 | 1.5 | N23W25 | C1.1 | 22:12 | 1025 (1543) | H | -5.3 | Dk | 10/02 01:25 | 7 |

[a]Proton intensity (Ip) in the >10 MeV channel; [b]Ip in the >50 MeV channel; [c]Soft X-ray flare size in the GOES 1–8Å channel; The C1.1 flare on 2013 September 29 was not listed, so we estimated it from the light curve. [d]First-appearance time of the CME in the LASCO C2 field of view. [e]Linear speeds obtained from the height-time plots with the space speed given in parentheses. [f]metric (m), decameter-hectometric (D) and kilometric (k). mk indicates type II bursts at all domains (metric to kilometric). Dk indicates type II bursts with components in the DH and kilometric domains, but not in the metric domain. [g]References that have information on these events: 1. Gopalswamy et al. (2002); 2. Gopalswamy et al. (2004); 3. Gopalswamy et al. (2005); 4. Vrsnak et al. (2003); 5. Gopalswamy and Yashiro 2013; 6. Gopalswamy (2012); 7. Gopalswamy et al. (2014).

Table 2. Filament and CME kinematics in the FE-SEP events

| Date | $a_i(F)$[a] m/s$^2$ | $a_i$ (flare) m/s$^2$ | $V_{EP}$ (1.5 R$_S$) km/s | $V_{EP}$ (20 R$_S$) km/s | $V_{CME}$ (Ext.) km/s[d] |
|---|---|---|---|---|---|
| 2000/09/12 | 154 | 770 | 275 | 677 | 2489 |
| 2002/05/22 | 272 | 800 | 258 | 621 | 1375 |
| 2011/11/26 | 108 | 280 | ----[c] | 878 | 1024 |
| 2013/09/29 | 104[b] | 460 | 297 | 603 | 1084 |

[a]Initial acceleration of the eruptive filament from EUV images. [b]Likely underestimate because the acceleration is determined by the first two CME core points and the initial position of the filament. [c]The eruptive filament motion could not be measured from EUV observations. [d]From the quadratic equation in Fig. 6 (blue lines) determined at the instant when the prominence core reached 20 R$_S$.

Table 3. Drift rates and shock speeds from Wind/WAVES type II radio bursts

| Date | $f_s$ MHz[a] | $f_r$ MHz[b] | α | Drift rate MHz/s | r Rs | $V_{sh}$ km/s | $V_{CME}$ km/s |
|---|---|---|---|---|---|---|---|
| 2000/09/12 | 24 | 3.26 | 4 | $3.27 \times 10^{-3}$ | 5.0 | 1744 | 1946 |
|  |  | 0.7 | 2 | $1.11 \times 10^{-4}$ | 20.17 | 2224 | 1946 |
| 2002/05/22 | 0.5 | 0.36 | 2 | $6.52 \times 10^{-5}$ | 14.38 | 1811 | 1718 |
| 2011/11/26 | 9 | 3.4 | 4 | $1.1 \times 10^{-3}$ | 7.13 | 802 | 935 |
|  |  | 0.6 | 3 | $6.9 \times 10^{-5}$ | 22.09 | 1178 | 1244 |
| 2013/09/29 | 13 | 5.75 | 2 | $1.88 \times 10^{-3}$ | 5.66 | 1287 | 1465 |
|  |  | 0.45 | 2 | $4.76 \times 10^{-5}$ | 23.0 | 1692 | 1596 |

[a]The starting frequency of type II bursts; [b]the reference frequency at which the drift rate and shock speed were estimated.

Table 4. List of large SEP events associated with C-class flares

| S. No. | Date | Ip (pfu) | γ | $V_{CME}$ km/s | W[a] | X-ray Imp | Location | Type II[b] |
|---|---|---|---|---|---|---|---|---|
| 1 | 2000/04/04 | 55 | 4.76 | 1188 | H | C9.7 | N25W55 | yY |
| 2 | 2000/10/25 | 15 | 3.82 | 770 | H | C4.0 | N10W66 | nY |
| 3 | 2004/04/11 | 35 | 4.01 | 1645 | 314 | C9.6 | S14W49 | nY |
| 4 | 2010/08/14 | 14 | 3.38 | 1205 | H | C4.4 | N17W52 | yY |
| 5 | 2012/09/28 | 28 | 3.25 | 1035 | H | C3.7 | N06W34 | nY |

[a]CME width in degrees; H denotes that the CME is a halo, so its true width is unknown. [b]the lowercase letters (y,n) denote whether or not a metric type II burst was associated with the SEP event (y=yes; n=no); the upper case letters provide similar information for DH type II bursts

Table 5. List of low-intensity (Ip ≤30 pfu) SEP events[a]

| S. No. | Date | Ip (pfu) | γ | $V_{CME}$ km/s | W | X-ray Imp | Location | Type II |
|---|---|---|---|---|---|---|---|---|
| 1 | 2000/07/22 | 17 | 2.53 | 1230 | 229 | M3.7 | N14W26 | yY |
| 2 | 2001/09/15 | 12 | 3.37 | 478 | 130 | M1.5 | S21W49 | yY |
| 3 | 2001/10/19 | 12 | 2.51 | 901 | H | X1.6 | N15W29 | yY |
| 4 | 2002/03/16 | 13 | 4.52 | 957 | H | M2.2 | S08W03 | nY |
| 5 | 2002/04/17 | 24 | 3.82 | 1240 | H | M2.6 | S14W34 | yY |
| 6 | 2002/08/14 | 26 | 4.58 | 1309 | 133 | M2.3 | N09W34 | yY |
| 7 | 2003/05/31 | 27 | 2.12 | 1835 | H | M9.3 | S07W65 | yY |
| 8 | 2011/08/09 | 26 | 1.74 | 1610 | H | X6.9 | N06W34 | yY |
| 9 | 2014/02/20 | 22 | 1.67 | 948 | H | M3.0 | S15W73 | yY |

[a]The columns have the same meaning as in Table 4.

**Figures**

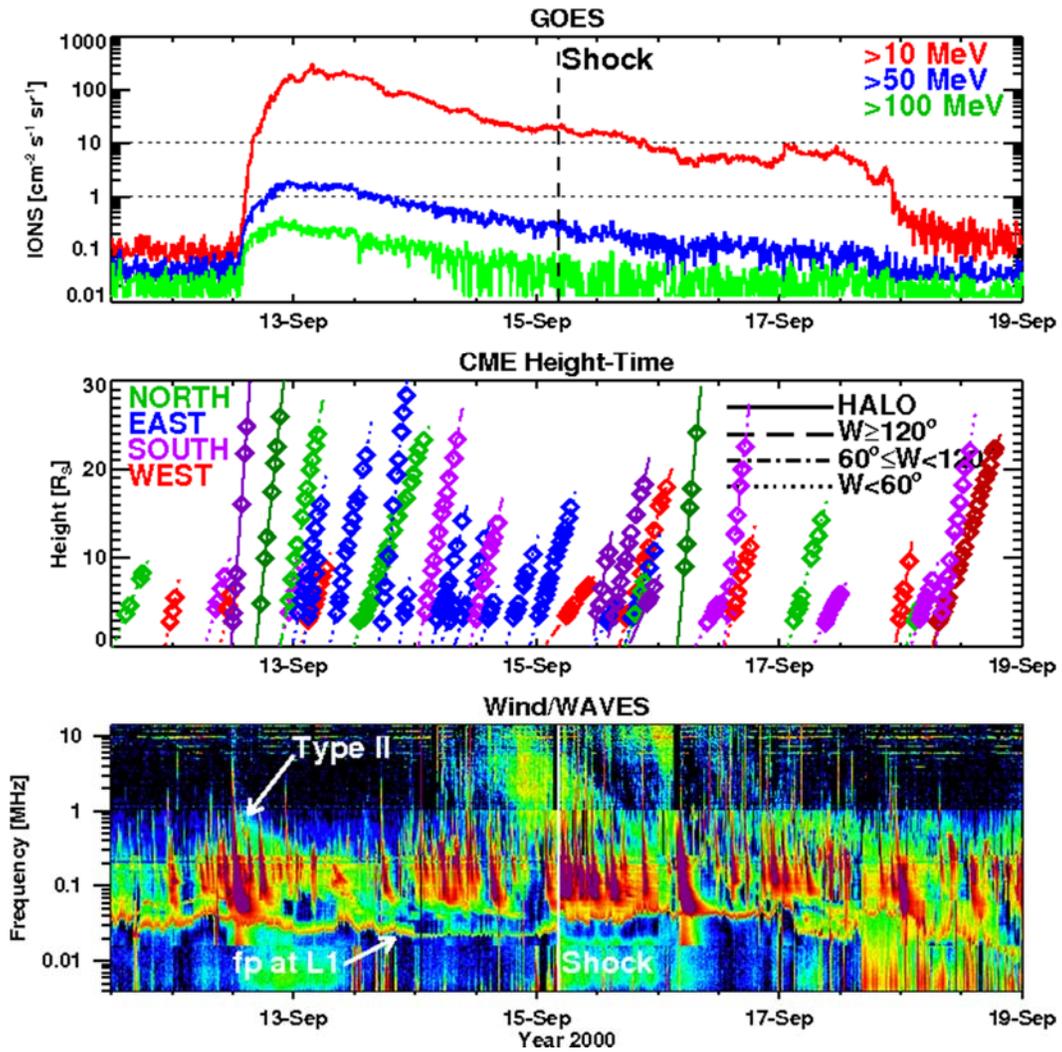

Figure 1a. GOES proton intensity (top), CME height-time plots (middle), and Wind/WAVES dynamic spectrum (bottom) for the 2000 September 12 SEP event. The vertical dashed line in the top panel marks the shock arrival at the SOHO spacecraft located at the Sun-Earth L1. The CME responsible for the SEP event is marked on the middle plot. The type II burst in the interplanetary medium and the shock arrival at the Wind spacecraft (marked by the vertical white line in the bottom panel) as a jump in the local plasma line (*fp*) are also marked.

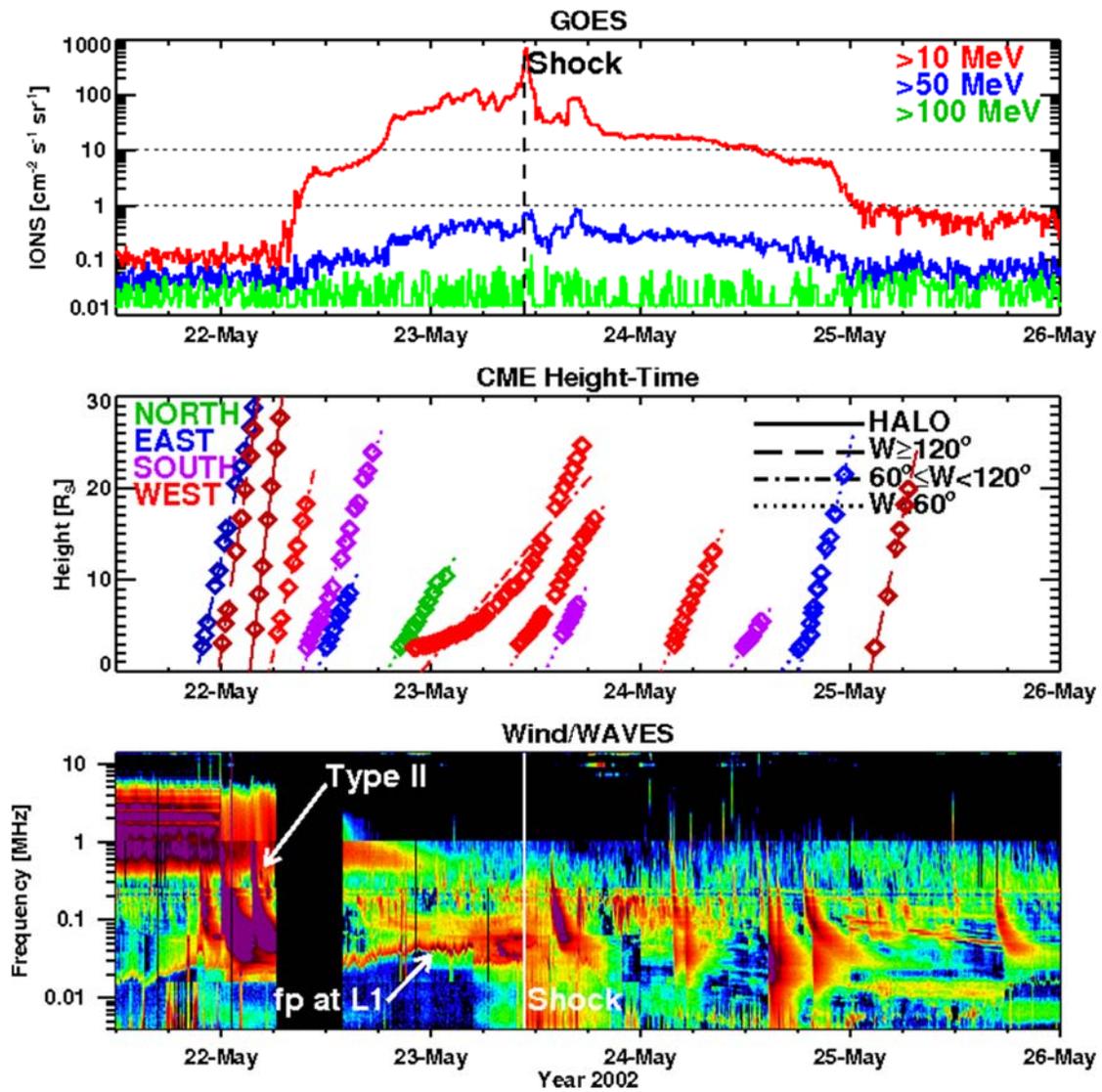

Figure 1b. Same as Fig.1a, but for the 2002 May 22 SEP event.

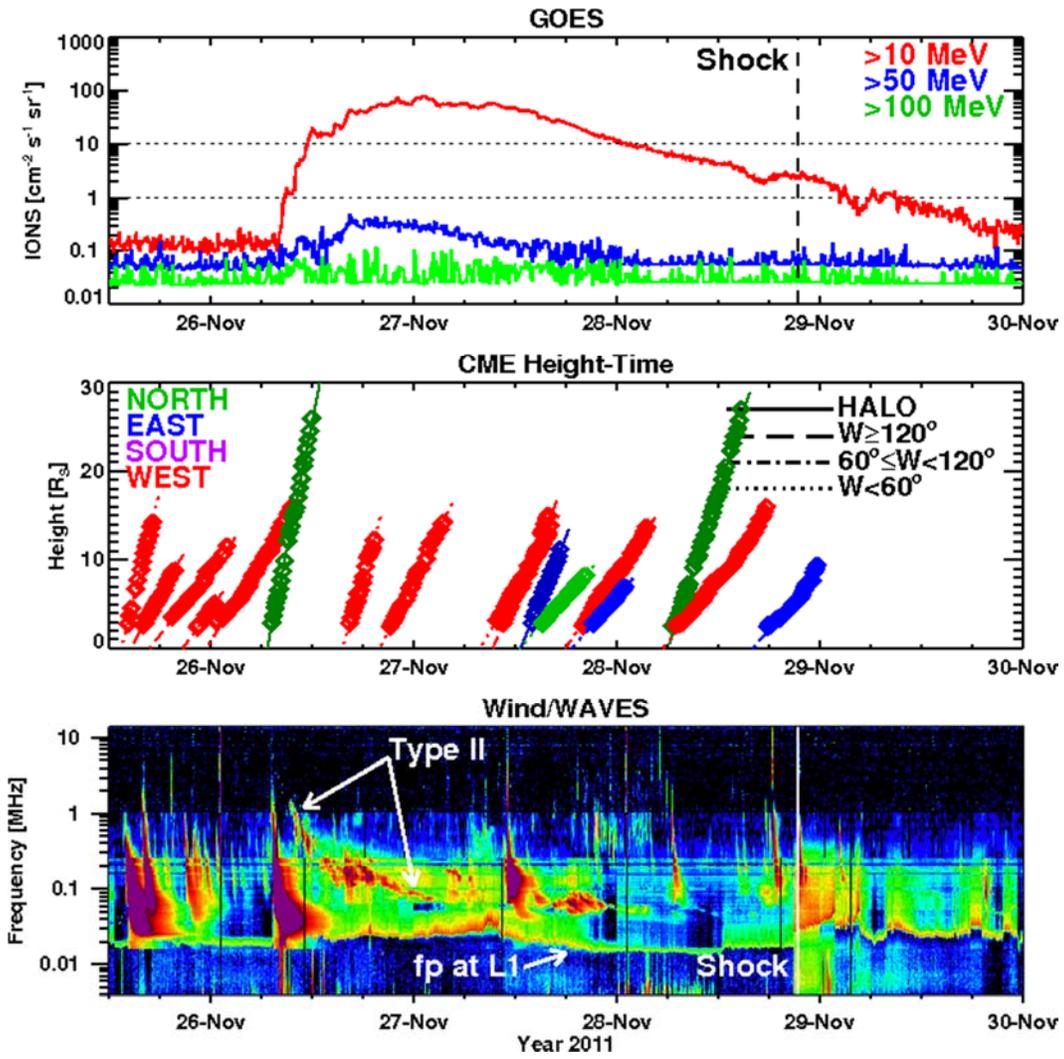

Figure 1c. Same as Fig.1a, but for the 2011 November 26 SEP event.

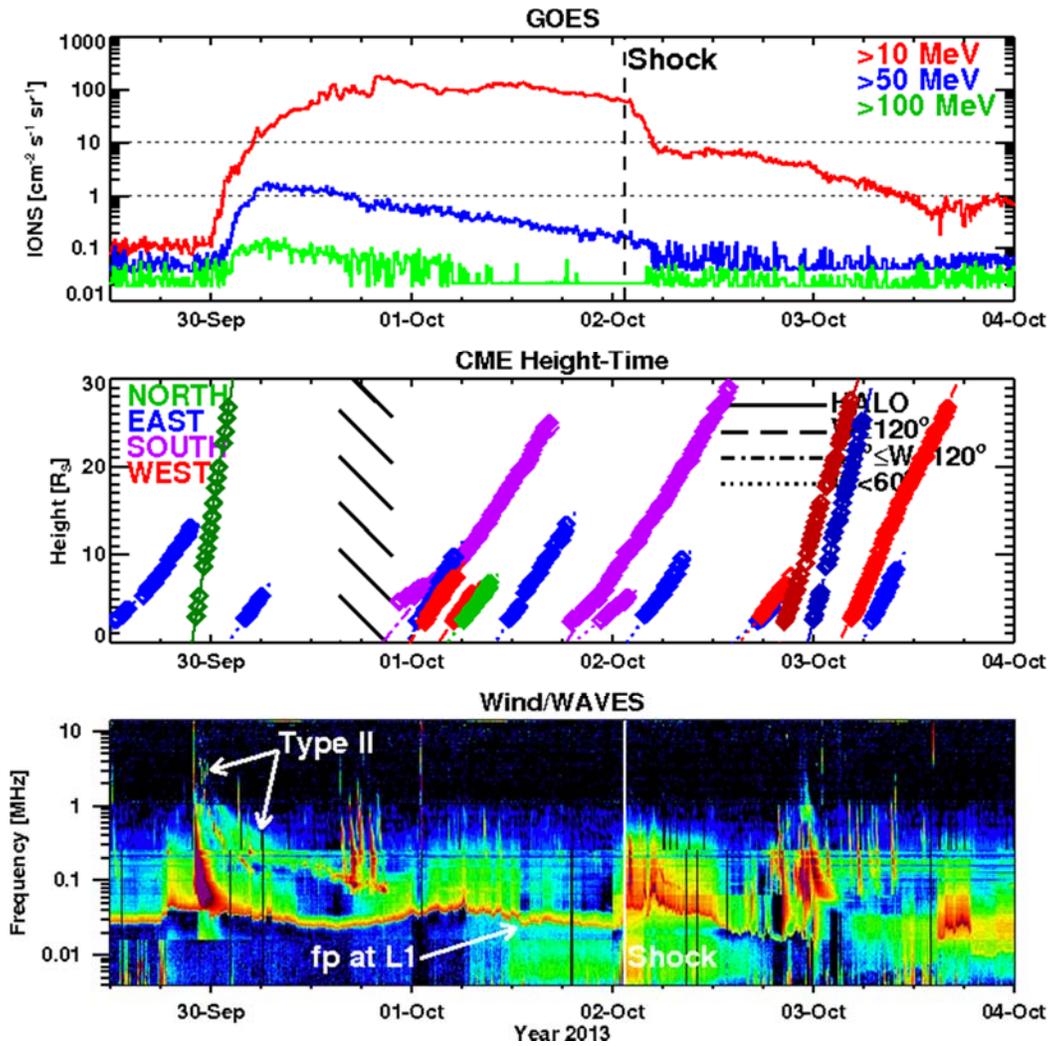

Figure 1d. Same as Fig. 1a, but for the 2013 September 29 SEP event. The hashed region in the height-time plots indicates a LASCO data gap.

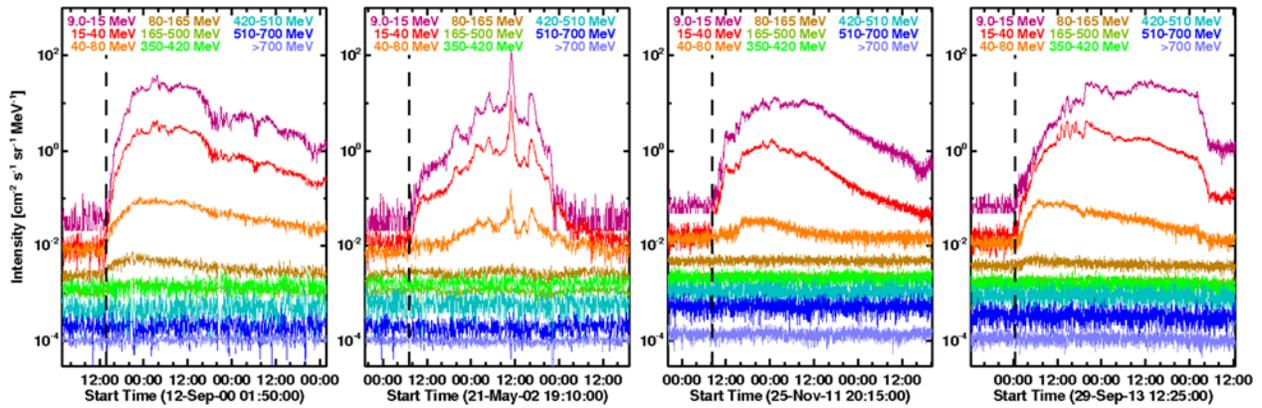

Figure 2. Proton fluxes for the four SEP events in Table 1. The vertical dashed line indicates the approximate onset times of the SEP events. The proton data are from National Geophysical Data Center (NGDC): http://satdat.ngdc.noaa.gov/sem/goes/data/new_avg/).

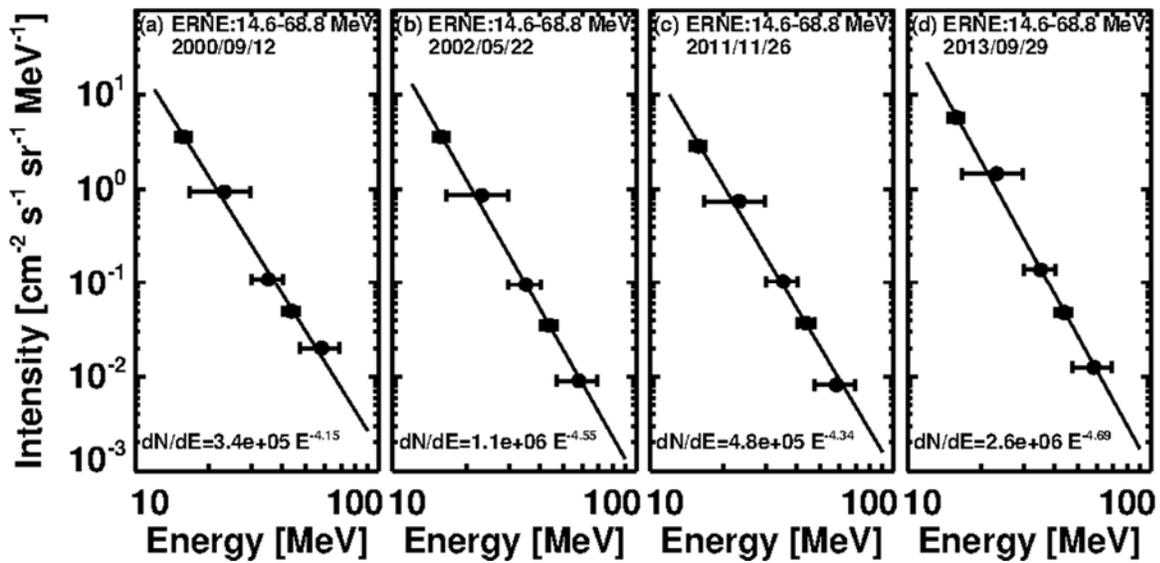

Figure 3. Time-of-maximum spectra for protons in the range 10 MeV <E<100 MeV obtained from the ERNE data for the four FE-SEP events. The effective energies in the channels are: 15.8, 23, 35.2, 43.8, and 58.2 MeV. The integration time is 1 hour in all energy channels. The horizontal bar on each data point corresponds to the width of the energy channel. The uncertainties in the proton fluxes are smaller than the sizes of the data points. The fitted spectra

have the form: $dN/dE = kE^{-\gamma}$ where $N$ and $E$ are the energetic particle density and energy, respectively, $k$ is a constant, and $\gamma$ is the spectral index.

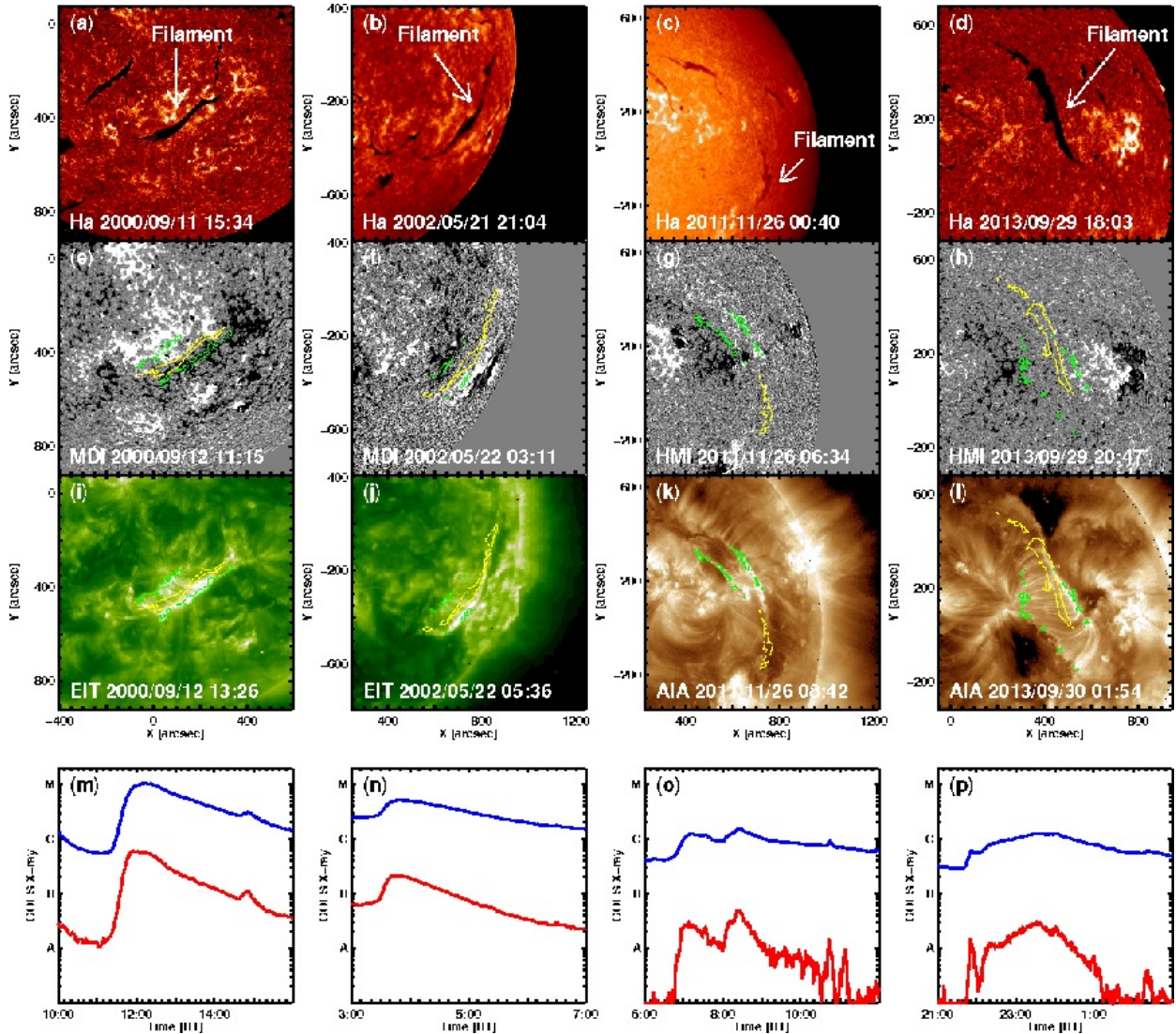

Figure 4. The pre-eruption filaments, flare ribbons, post-eruption arcades, and the GOES soft X-ray light curves for the four events. The H-alpha filaments in the top row were observed at the Big Bear Solar Observatory in the United States (a-b, d) and the Hida Observatory in Japan (c). The flare ribbons (green contours) and the original locations of the filaments (yellow contours) are superposed on photospheric magnetograms (SOHO/MDI for the cycle-23 events and SDO/AIA for the cycle-24 events). The post-eruption arcades were observed by SOHO/EIT (i,j) and SDO/AIA (k,l). The GOES plots in the bottom row show that the soft X-ray flares ranged from C1 to M1.

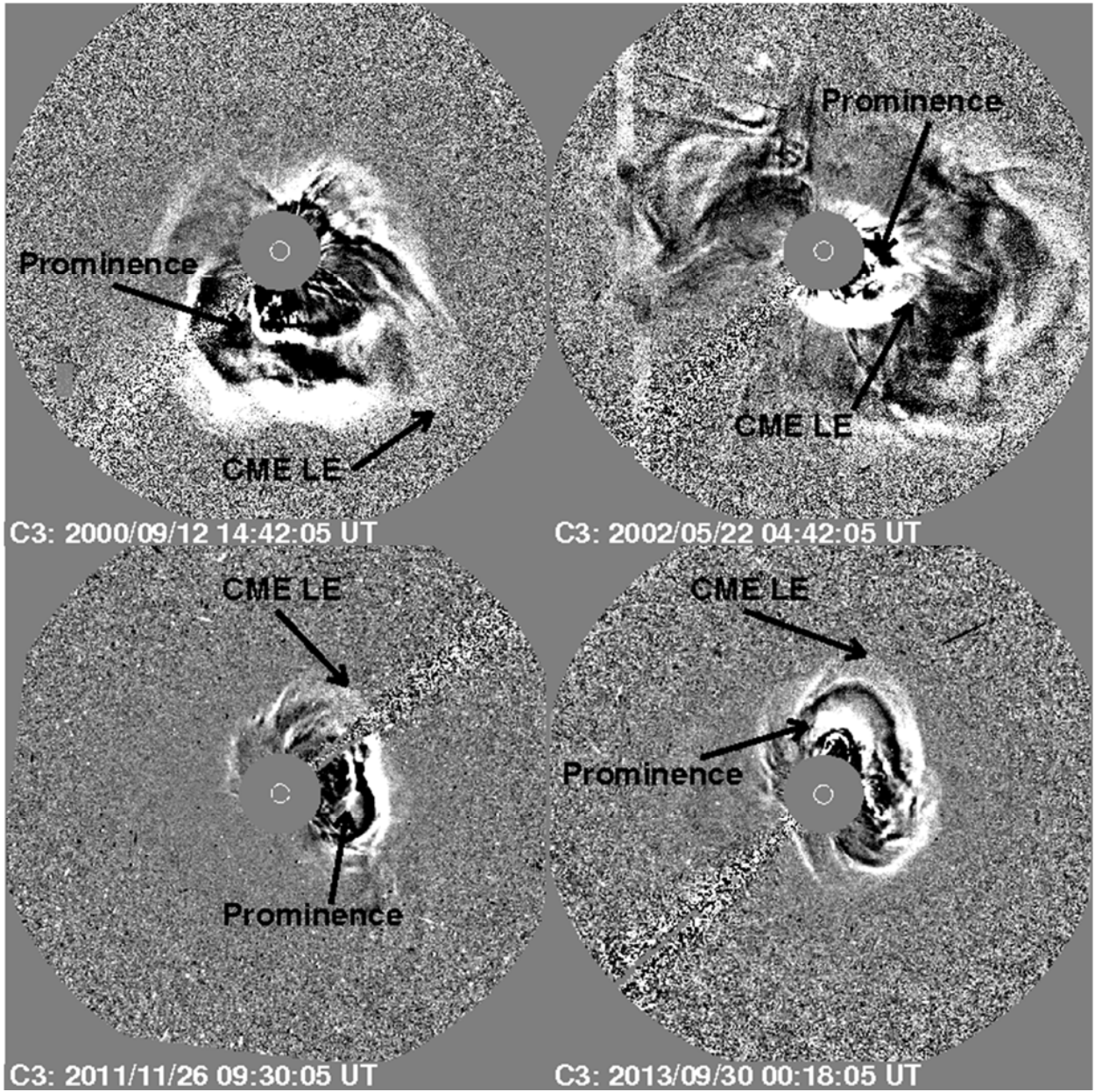

Figure 5. SOHO/LASCO images showing the CMEs and the associated eruptive prominences in the four SEP events. All four CMEs became halos in the field of view of the outer coronagraph (C3) of SOHO/LASCO.

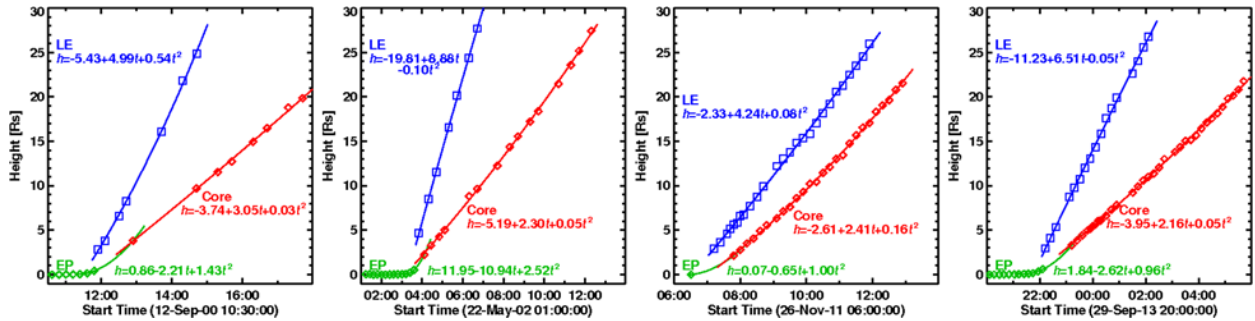

Figure 6. Height-time plots of the eruptive filaments and the associated CMEs for the four events. The height-time history of the eruptive prominences (EPs) measured from EUV images are shown in green. The green solid curves are second-order fits to the EP height-time measurements from EUV and one prominence core data point from LASCO. The EPs that became prominence cores in the LASCO images (see Fig. 5) are also measured and fitted to a different curve (solid curve and diamond symbols in red). The height-time history of the CME leading edges (marked LE) are also shown in blue. The equations of the fitted curves are also shown on the plots (time in hours from the start time and the height is in solar radii measured from the Sun center).

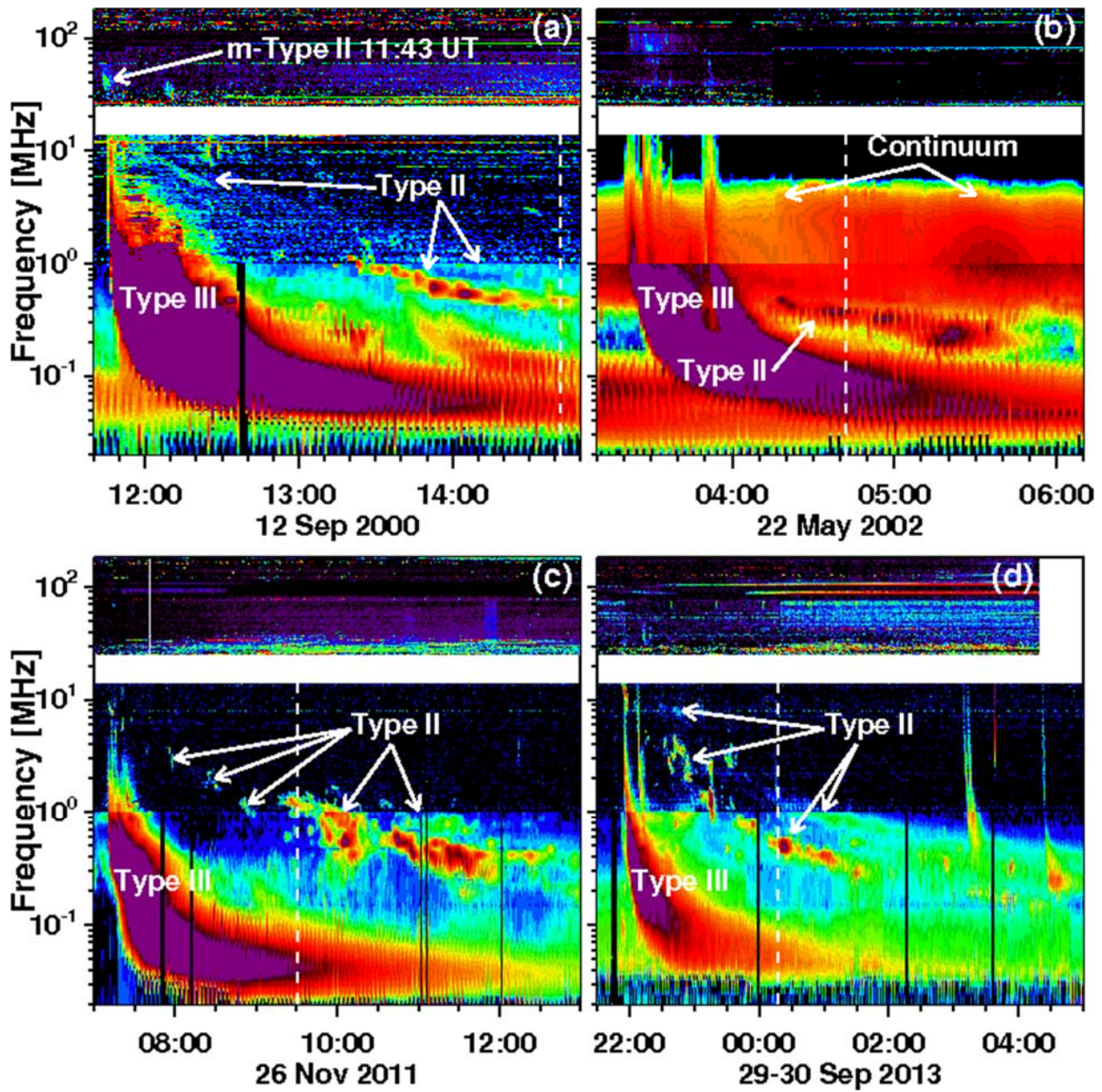

Figure 7. Dynamic spectra combining ground based observations with Wind/WAVES data showing type II and type III radio bursts. Fundamental-harmonic structure can be seen in all but the 2002 May 22 event. In this event, there was a long-lasting continuum, which did not allow us to determine whether a harmonic component was present or not. The vertical dashed lines mark the times of the LASCO images shown in Fig.5.

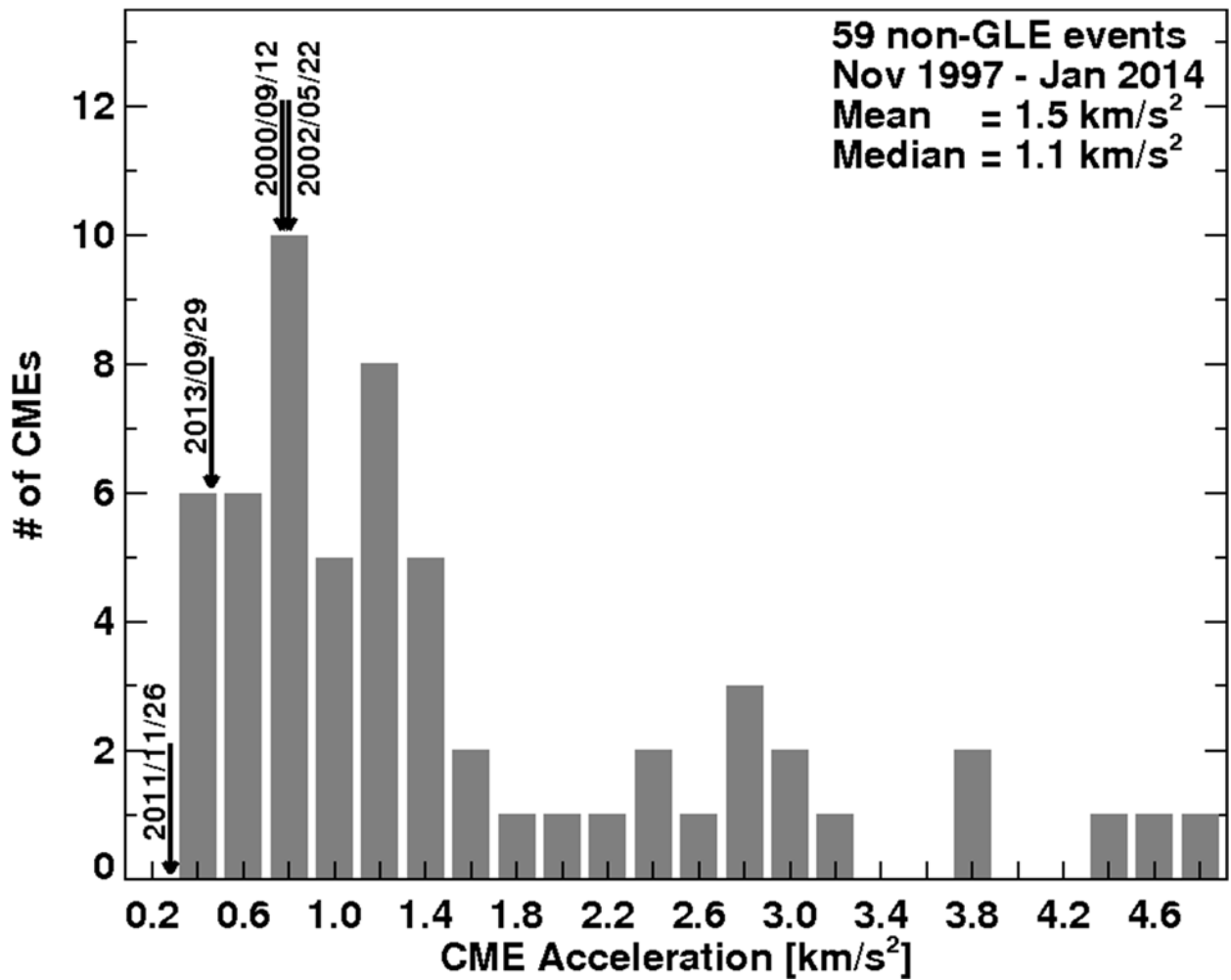

Figure 8. CME accelerations derived from the flare rise time and the CME speed in the coronagraph FOV. The mean and median values of the distribution are marked on the plot. The acceleration values of the four FE-SEP events are shown by arrows.

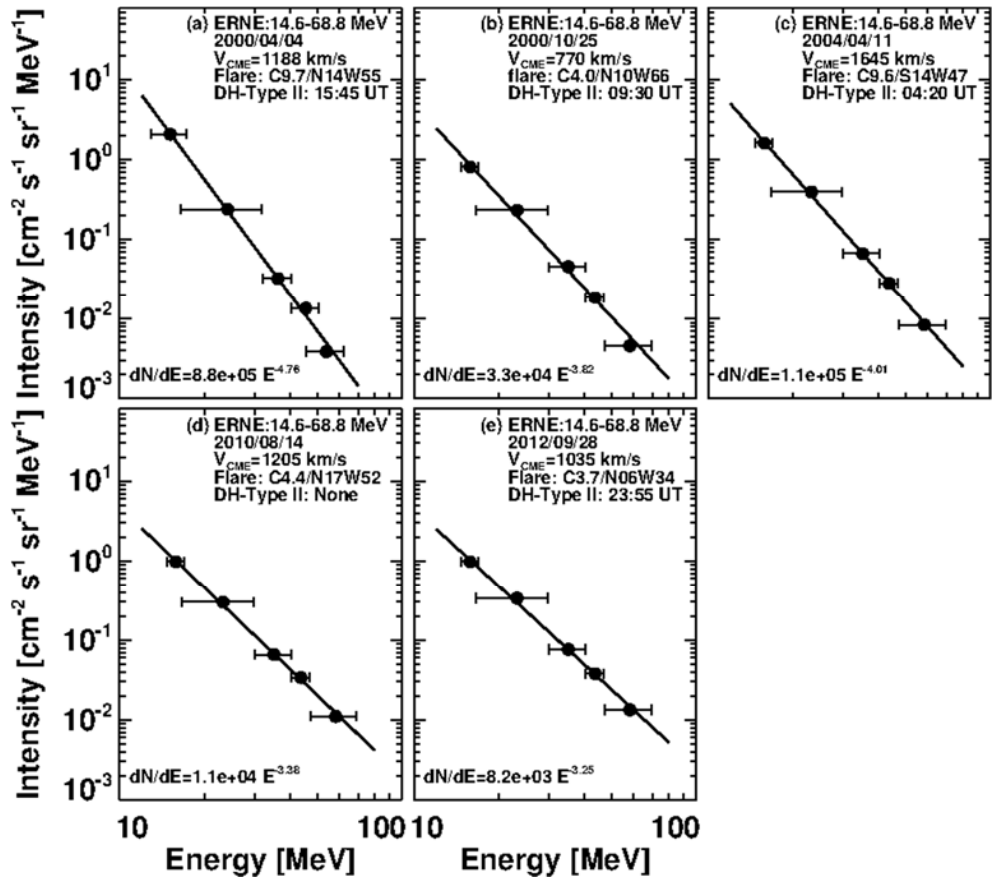

Figure 9. Time-of-maximum SEP spectra for the five large SEP events associated with C-class flares. The integration time is 1 hour in all energy channels.

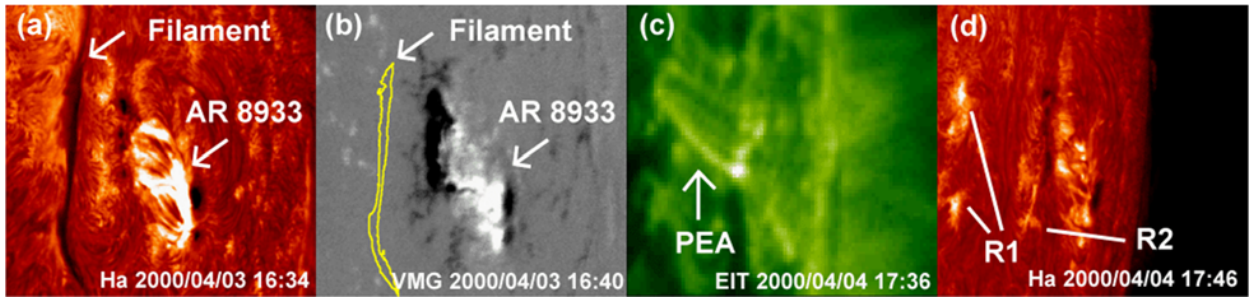

Figure 10. The source region of the 2004 April 4 SEP event. (a) BBSO H-alpha image taken on 2004 April 3 at 16:34 UT showing the pre-eruption stage of the filament and AR 8933 to its west. (b) BBSO video magnetogram (VMG) taken a few minutes after the H-alpha image; the position of the H-alpha filament is shown on the magnetogram (yellow contour). (c) SOHO/EIT image obtained on April 4 at 17:36 UT showing the post eruption arcade. (d) H-alpha image taken in the post- eruption stage (2004 April 4 at 17:46 UT showing the two-ribbon structure to the east of the active region.